\documentclass[twocolumn,showpacs,preprintnumbers,amsmath,amssymb]{revtex4}

\usepackage{graphicx}
\usepackage{bm}

\newcommand{\beq}{\begin{equation}}
\newcommand{\eeq}{\end{equation}}
\newcommand{\bea}{\begin{eqnarray}}
\newcommand{\eea}{\end{eqnarray}}

\begin{document}


\title{Condensation in disordered lasers: theory, 3D+1 simulations and experiments}

\author{ C. Conti$^{1}$, M. Leonetti$^{2}$, A. Fratalocchi$^{3,1}$, L. Angelani$^{4}$, G. Ruocco$^{1,2}$}

\affiliation{
$^1$Research center Soft INFM-CNR, c/o Universit\`a di
Roma ``Sapienza,'' I-00185, Roma, Italy \\
$^2$Dipartimento di Fisica, Universit\`a di Roma ``Sapienza,''
I-00185, Roma, Italy \\
$^3$Centro studi e ricerche ``Enrico Fermi,'' Via Panisperna 89/A,
I-00184, Roma, Italy   \\
$^4$Research center SMC INFM-CNR, c/o Universit\`a
di Roma ``Sapienza,'' I-00185, Roma, Italy
}

\date{\today}

\begin{abstract}
The complex processes underlying the generation of a coherent-like emission 
from the multiple-scattering of photons and wave-localization in the presence
of structural disorder are still mostly un-explored. 
Here we show that a single nonlinear Schroedinger equation,
playing the role of the Schawlow-Townes law for standard lasers, quantitatively reproduces
experimental results and three-dimensional time-domain parallel simulations of a colloidal
laser system.
\end{abstract}


\maketitle

\begin{figure}
\includegraphics[width=8.5cm]{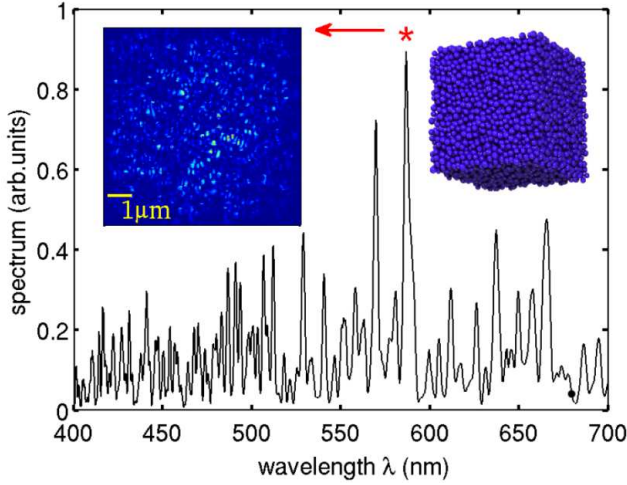}
\caption{
\label{figsc}(Color online) Electromagnetic spectrum as obtained after a wide-band excitation; 
the left inset shows the energy density in the middle section ($\lambda=586.5$~nm, asterisk);
a sketch of the system is also shown on the right.
}
\end{figure}

\begin{figure}
\includegraphics[width=8.5cm]{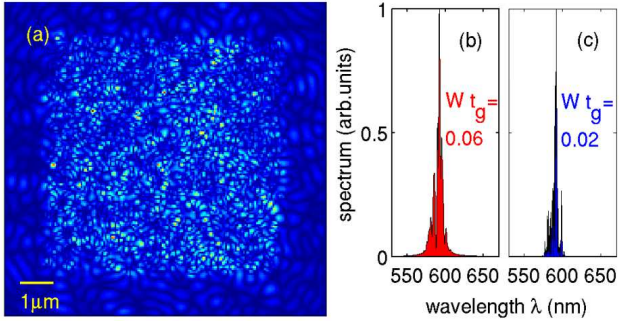}
\caption{\label{figprofilesims}
(Color online) 3D+1 Maxwell-Bloch simulation of RL:
(a) energy distribution in the sample middle section ($N_a\times10^{24}m^{-3}$);
(b) spectrum for $N_a=2\times10^{24}m^{-3}$; (c) as in (b) for $N_a=5\times10^{24}m^{-3}$.
}
\end{figure}

\begin{figure}
\includegraphics[width=8.5cm]{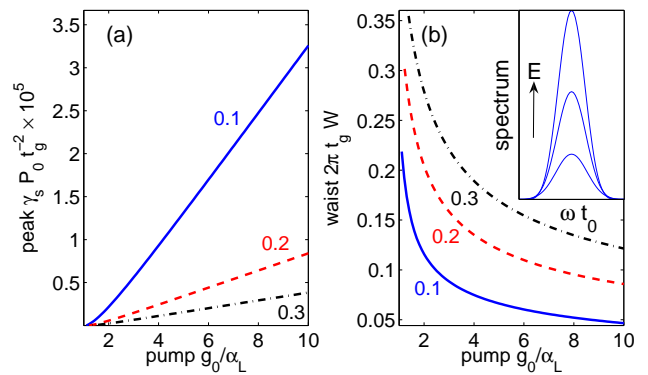}
\caption{
\label{figfeat} (Color online) Theory:
Peak spectrum (a) and spectral waist (b) for various $\kappa$ versus the pumping rate in dimensionless units.
Inset: Spectral profiles for $E=1.1,10,20$.
}
\end{figure}

\begin{figure}
\includegraphics[width=8.5cm]{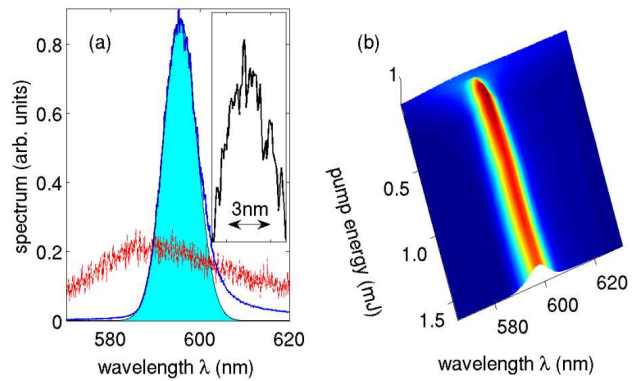}
\caption{\label{fig4} (Color online)
Experimental results: (a) spectra at energies $20~\mu$J (thin line) and $1000~\mu$J,
(shaded area is a Gaussian fit), inset: corresponding enlarged central spectral region;
(b) unitary-area averaged spectra ($100$ shots) Vs energy.
}
\end{figure}

\begin{figure}
\includegraphics[width=8.5cm]{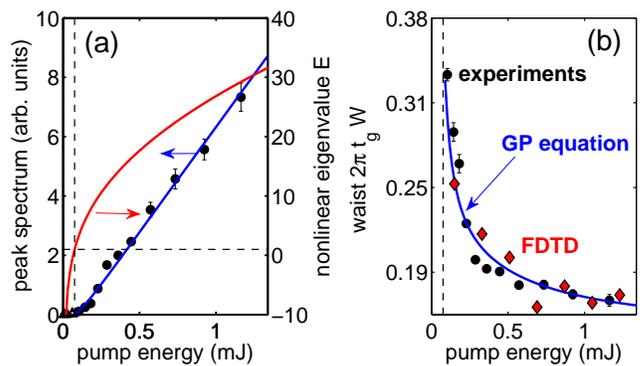}
\caption{\label{fig5} (Color online)
(a)  Left scale, experimentally retrieved laser peak spectrum versus pump energy, 
the line is a best fit from theory; right scale, nonlinear eigenvalue 
(dashed lines corresponds to $E=1$). (b) Linewidth Vs energy; the continuous line is
the best fit from theory; the FDTD Maxwell-Bloch simulations (diamonds) are also shown.
}
\end{figure}

Random lasers (RL) are  a rapidly growing field of research, with 
implications in soft-matter physics,
light localization and photonic devices \cite{Cao05,Wiersma:08}.
Since the pioneering investigations \cite{Ambartsumyan:66,Lethokov68},
different groups reported on experimental observations, from 
paint pigments to human tissue \cite{Lawandy94,Siddique:95,Zhang:95,Polson04,vandermolen:06b}.
In all of these cases a coherent-like narrow spectral line emerges from the
fluorescence as the pump energy is increased
and, in some instances, several spectral peaks have been reported \cite{vandermolen:06b,Mujumdar:07}.
\\ In standard single-mode lasers, without structural disorder, 
the emission linewidth is linked to the electromagnetic energy
stored in the cavity by the so-called Schawlow-Townes (ST) law \cite{YarivBook,Schawlow:58}.
An equivalent law for RL is missing.
Nevertheless various issues (like the statistical properties 
and the link with spin-glass theory \cite{Beenakker:98,Patra02,Hackenbroich:01,vandermolen:06b,Cao03r,Angelani06}),
were theoretically analysed, 
while the leading model (quantitatively compared with experiments) is that 
based on the light-diffusion approximation \cite{Wiersma96,John96,Florescu04,Lubatsch05},
which however overlooks the ondulatory character of the involved photons.
Within a different perspective, RL is due to several localized electromagnetic (EM) states put into oscillations
in a disordered environment (as, e.g., in \cite{Angelani06,vandermolen:06b,Deych05,Tureci:08}). 
In this framework, it is expected that
the number of involved modes increases with the pump energy and, correspondingly, the spectrum widens. 
However, exactly the opposite happens and this is also accompanied by the shortening of the emitted pulse 
\cite{Goeuedard:93,Siddique:96, Papadakis:07}.
In addition, the fact that strong (or Anderson) localization of light sustains the 
RL action is still debated. Ab-initio computational studies were limited
to 1D and 2D geometries \cite{Jiang00,Vanneste02}, not accounting for the critical character of three-dimensional (3D) localization \cite{ShengBook}.
Monte-Carlo simulations neglect interference effects \cite{Balachandran:97,Berger:97,Mujumdar04}.
\\Here we report on an original theoretical formulation; we quantitatively compare its predictions with experiments 
and with the first ever reported 3D+1 ab-initio Maxwell-Bloch simulations.
We show that the RL linewidth is ruled by a nonlinear differential-equation, which is the equivalent of the ST-law,
and is formally identical to the 
nonlinear Schroedinger, or Gross-Pitaevskii (GP), equation governing ultra-cold
atoms \cite{Dalfovo:99}. There is hence a strict connection between photons in RL
and ultra-cold bosons;
the spectral narrowing observed in RL is thus ascribed 
to a {\it condensation process} \cite{connaughton:263901} of the involved electromagnetic resonances.
\\\noindent{\it Simulations ---}
We consider a vectorial formulation of the Maxwell-Bloch (MB) equations \cite{Conti07Mie,Conti08PhC}. 
$21$ nonlinear stochastic partial differential equations are solved by finite-difference
time-domain (FDTD) discretization on a grid distributed on (typically) $256$ processors.
We model an active medium that is infiltrated in the voids of a granular distribution of particles obtained by molecular dynamics \cite{Conti07}. 
We consider $8000$ TiO$_2$ particles (average diameter $300$~nm) with refractive index $2.9$ (Fig. \ref{figsc}). 
The gain bandwidth is $230$~nm ($\sim~1/t_g$ with $t_g$ the life-time)
and the central wavelength is $\lambda_0=590$~nm ($\omega_0$ is the angular frequency).
Amplification is only present in the interstices between colloidal spheres 
with pump-rate (varied by the atomic inversion density $N_a$, see  \cite{Conti07Mie,Conti08PhC}) constant over the
$\sim3$~ps simulation.
The lasing action is self-starting from the noise due spontaneous emission (SE) modeled as a stochastic term.
\noindent 
\\In absence of light-amplification, the response to a single cycle pulse ($\sim 1$~fs) at wavelength $\lambda=532$~nm,
(Fig. \ref{figsc}) unveils several spectral peaks corresponding to long-living modes.
Field spatial distribution (inset of Fig.\ref{figsc}) is determined by continuous-wave (CW) excitation.
\\ \noindent Then we simulate the RL action:
when increasing the pumping a coherent field is built from noise.
Fig.\ref{figprofilesims}a shows the snapshot of the EM energy density in the
sample middle section. In Fig.\ref{figprofilesims}b,c we display the spectra for two pumping levels
in quantitative agreement with the experiments below.  
The 3D-RL action is mediated by several modes with overlapping resonances.
In real-world samples the RL volume and the number of modes $N$ is much larger than that found in
our simulations; the outcome is a smoother emission profile (Fig.\ref{fig4}).
\\\noindent{\it Gross-Pitaevskii equation ---}
The RL frequency content is given by $A(\omega)$, such that
$|A(\omega)|^2$ is the energy stored in the disordered cavity 
at relative angular frequency $\omega$ with respect to $\omega_0$. 
The loss coefficient is $\alpha(\omega)$, and the gain $g$
depends on the whole shape $A(\omega)$, as due to the nonlinear susceptibility
of the resonant medium \cite{Angelani06}.
In the frequency domain, the oscillation condition ``gain$=$loss'' reads as
$g[A(\omega)]=\alpha(\omega)A(\omega)$.
Limiting for the moment to the loss profile $\alpha(\omega)$,
and following the previous numerical analysis, one has that,
in a disordered system sustaining various resonances, $\alpha(\omega)$
is a smooth function interleaved by narrow spectral dips,
corresponding to $N$ localized (high Q-factor) long-living modes (see figure \ref{figsc}).
In this case the loss function $\alpha(\omega)$ can be modeled as
\beq
\alpha(\omega)=\alpha_0(\omega)-\sum_{j=1}^{N}\alpha_j(\omega-\omega_j)\text{,}
\label{l1}
\eeq
where $\alpha_0(\omega)$ is the non-resonant smooth loss profile and $\alpha_j(\omega-\omega_j)$ is a sharply peaked (centered at $\omega_j$) lineshape
corresponding to the localized mode $j$ 
[$\alpha_j(\omega)$ is centered at $\omega=0$ for later convenience].
Since the RL spectral line is limited, we take $\alpha_0(\omega)\cong\alpha_0$. 
For large systems, we also expect a huge number of modes \cite{vandermolen:06b}
with comparable properties,
we will hence take $\alpha_j(\omega)\cong \alpha_{avg}(\omega)$, where $\alpha_{avg}$ is an average resonant line-shape. 
Therefore the oscillation condition becomes
\beq
\label{o1}
\begin{array}{l}
\displaystyle g[A(\omega)]=\alpha_0(\omega) A(\omega)-\sum_j \alpha_j(\omega-\omega_j)A(\omega)\cong\\
\displaystyle \cong \alpha_0 A(\omega)-\sum_j \alpha_{avg}(\omega-\omega_j)A(\omega_j)\text{,}
\end{array}
\eeq
where we exploited the fact than $\alpha_j(\omega-\omega_j)$ is much narrower than $A(\omega)$,
and hence it ``samples'' the emission spectrum at $\omega_j$.
In the continuum limit the right-hand side of Eq. (\ref{o1}) becomes
\beq
\alpha(\omega) A(\omega)\cong \alpha_0 A(\omega)- \int \alpha_{avg}(\omega-\Omega)A(\Omega)d\Omega\text{.}
\label{o2}
\eeq
We then consider the left-hand side (amplifying part) $g[A(\omega)]$ of Eq. (\ref{o1}) and
we exploit the passive mode-locking laser theory \cite{Haus:00,Kutz:06}.
For a finite gain bandwidth with life-time $t_g$, Eq.~(\ref{o1}) in the time-domain is
\beq
\label{g3}
 g_0 \left[ a(t) + t_g^2 \frac{d^2 a}{dt^2}- \gamma_s |a|^2 a\right] =\left[ \alpha_0-\phi_L(t)\right] a\text{,}
\eeq
where we introduced 
the Fourier transform $a(t)$ of $A(\omega)=\mathcal{F}[a]=(1/2\pi)\int a(t) \exp(i \omega t) dt$ and
 $\phi_L=\mathcal{F}[\alpha_{avg}]$.
In Eq.~(\ref{g3}) $g_0$ is the small signal gain and $\gamma_s$ is the gain saturation coefficient  \cite{Haus:00,Kutz:06}.
$\alpha_{avg}(\omega)$ is narrow with respect to the
gain bandwidth, hence $\phi_L(t)$ can be expanded around $t=0$: 
$\phi_L(t)\cong (\alpha_0-\alpha_L) \left[1- (t/t_L)^2\right]$, 
where $\alpha_L$ is the average loss for the high-Q modes ($\alpha_0>\alpha_L$) and
$t_L$ is their average lifetime.
Equation (\ref{g3}) is then cast in a dimensionless form 
$a=a_0 \varphi(\tau)$ and $t=\tau t_0$ with 
$a_0^2=t_g \sqrt{\alpha_0-\alpha_L}/\gamma_S \sqrt{g_0} t_L$ and $t_0^2=t_g t_L\sqrt{g_0}/\sqrt{\alpha_0-\alpha_L}$:
\beq
\label{g5}
-\frac{d^2 \varphi}{d\tau^2}+\tau^2 \varphi+|\varphi|^2 \varphi=E \varphi\text{.}
\eeq
The ``nonlinear eigenvalue'' $E$ is given by
\beq
\label{ee}
E=\frac{t_L}{t_g}\frac{g_0-\alpha_L}{ \sqrt{(\alpha_0-\alpha_L)g_0}}=\frac{p-1}{\kappa \sqrt{p}}\text{,}
\eeq
where $p=g_0/\alpha_L$ is proportional to the pump energy, and
$\kappa\equiv(t_g/t_L)(\alpha_0/\alpha_L-1)^{1/2}$.
\\ 
It is known that wave-resonances in random systems display 
a distribution of decay times that is bell-shaped around some value $t_L$ and with comparable width \cite{Chabanov03}.
Equation (\ref{g5}) is the oscillation condition for these modes with different 
$\tau$ (which corresponds to the shift from $t_L$),
including gain saturation, finite gain bandwidth and the mode-coupling due to the overlapping resonances.
Equation (\ref{g5}) [or Eq.(\ref{g3})] is identical to the bound state GP equation for the
1D Bose-Einstein condensation with an external potential $\phi_L(t)$.
This shows that a spectral region of high Q-modes acts as a
trapping potential for the energy levels of the excited photons.
Frequencies tend to be concentrated in this spectral range,
as Bose-condensed atoms tend to be localized by the external trap \cite{Dalfovo:99}.
\noindent Equation (\ref{g5}) displays bell-shaped solutions for $E>1$ (see, e.g., \cite{Kivshar:99}), this
implies a pumping threshold for the laser action;
the corresponding dimensionless gain 
$p_{th}\cong1+\kappa^2/2+\kappa\sqrt{4+\kappa^2}/2$ is given by $E=1$ 
($p_{th}\cong 1$ as $\kappa<<1$).
As $E\gtrsim 1$ ($p\gtrsim p_{th}$,), an approximated solution of Eq. (\ref{ee}) is 
$\varphi=2^{1/4}\sqrt{E-1}\exp(-\tau^2/2)$. The RL spectrum at the threshold (i.e. for $p\sim p_{th}$) is hence
\begin{equation}
S(\omega)=|A(\omega)|^2=\frac{t_g^2}{\sqrt{2}\pi\gamma_S} (E-1) \exp\left[-\frac{\omega^2}{8\pi^2 W_{th}^2}\right]\text{,}
\label{s1}
\end{equation}
with the spectral waist (in frequency $\omega/2\pi$)
\begin{equation}
2\pi t_g W_{th}=\sqrt{\frac{\kappa}{2}}=\sqrt{\frac{t_g}{2t_L}\sqrt{\frac{\alpha_0}{\alpha_L}-1}}.
\label{wth}
\end{equation}
Eq. (\ref{wth}) implies that the RL linewidth at threshold 
is a fraction of the gain bandwidth ($\sim 1/t_g$) given by $\sqrt{\kappa}/2 \sqrt{2} \pi<<1$.
For a gain bandwidth of $250~$nm and a RL spike linewidth $\lessapprox 0.5$~nm
(i.e. $t_L/t_g\cong500$) and taking for the losses $\alpha_L\cong \alpha_0/1000$
\cite{nota},
it is $\kappa\cong0.1$ and $2\pi t_g W_{th}\cong0.2$.
For $E>1$, the spectral profile 
is obtained by the numerical solution of (\ref{g5}); from its Fourier transform $\tilde \varphi(\omega t_0)$
(inset in Fig.~\ref{figfeat}b) 
the normalized waist $w_\varphi[E(p)]$ and peak $p_\varphi[E(p)]$ are determined;
$W=w_\varphi \sqrt{\kappa}/ p^{1/4}t_g$ 
and $S_p=t_g^2 p_\varphi/\gamma_S$ are the corresponding for $S(\omega)$ in real-world units (Figs. \ref{figfeat} and \ref{fig5}).
\\Summarizing, Eq.(\ref{ee}) relates the pumping $p$ to the 
nonlinear eigenvalue $E$, which fixes the spectral line-shape through Eq.(\ref{g3}) [or Eq.(\ref{g5})];
this equation can be hence considered as the equivalent for RL of the ST-law.
\\{\it Experimental results ---} 
We use a colloidal dispersion of TiO$_2$ (Sachtleben Hombitan R611) particles in methanol
doped by Rhodamine B (Sigma-Aldrich R6626); the packing fraction is $0.2$ with average index $\bar n=1.5$
(the measured mean free path by enhanced back-scattering for pure methanol is $\ell=1700$~nm at $\lambda=532$~nm);
the RL pump is a $120$~ps linearly polarized $10$~Hz Nd:Yag laser at $532$~nm and $0.8$~mm spot-size.
Emission is retrieved by a fiber coupled spectrograph (Jobin Yvon, focal length $140$~mm) and
a thermoelectrically cooled CCD camera.
Figure \ref{fig4} shows the width (standard deviation) and the peak of the spectrum
averaged over $100$ laser shots versus pump energy $\mathcal{E}$;
the RL line first narrows and then stabilizes to a smooth profile.
The best-fit with the theory (Fig. \ref{fig5}) furnishes  $\kappa\cong 0.14$; 
the threshold pump energy ($E=1$) is $\mathcal{E}_{th}\cong 0.09$~mJ.
In Fig.~\ref{fig5}b we also display the linewidth calculated by a Gaussian fit of the FDTD data 
(the energy axis has been scaled to fit the experiments).
\\{\it Conclusions ---}
A theoretical approach based on a nonlinear bound-state equation,
identical to the GP equation for BEC, \cite{Dalfovo:99} has been shown
to quantitatively agree with experimentally retrieved laser emission in a colloidal dye-doped dispersion of TiO$_2$ particles
and with 3D+1 first-principle numerical simulations. RL emission can be related to a condensation process of several wave-resonances in the presence
of disorder; the distribution of their decay times playing the role of a temporal trapping potential. 
The simultaneous spectral and temporal narrowing with the number photons in RL
is hence corresponding to the spectral and spatial narrowing of the BEC wave-function
at the condensation.
\\{\it Acknowledgments ---}
We acknowledge support from the INFM-CINECA initiative for parallel computing.
The research leading to these results has received funding from the European Research Council
under the {\it European Community}'s Seventh Framework Program (FP7/2007-2013)/ERC {\it grant agreement} n.201766.

\begin{thebibliography}{42}
\bibitem[{not()}]{nota}
\bibinfo{note}{$1/\alpha_0$ is taken as the time needed to travel a mean free
  path $\ell$ at velocity $c/\bar n$; with $\bar n=1.5$ and $\ell\cong1700$~nm
  ($k\ell\cong20$), one finds $1/\alpha_0\cong10$~fs, and assuming
  $1/\alpha_L\cong2 t_L\cong 10$~ps (i.e. $\alpha_0/\alpha_L\cong1000$).}
\end{thebibliography}

\end{document}